\documentclass[aps,prb,twocolumn,showpacs,amsmath,amssymb,floatfix]{revtex4-1}
\usepackage{hyperref,graphicx,amsmath,amssymb,latexsym,amsfonts,dsfont,relsize,color,bm,cmap,lmodern}
\usepackage[utf8]{inputenc}
\usepackage[T1]{fontenc}
\usepackage[english]{babel}
\usepackage[caption=false]{subfig}
\usepackage[usenames,dvipsnames,svgnames,table]{xcolor}
\definecolor{rltred}{rgb}{0.75,0,0}
\definecolor{rltgreen}{rgb}{0,0.5,0}
\definecolor{rltblue}{rgb}{0,0,0.75}
\hypersetup{colorlinks,%
hypertexnames=true,%
pdfauthor={Adrian Girschik, Florian Libisch,
  Stefan Rotter},%
pdftitle={Percolating States in the Topological Anderson Insulator},%
pdfview=FitH,%
pdfstartview=FitH,%
pdfpagemode=UseNone,%
bookmarksopen=true,%
bookmarksnumbered=true,%
pdfhighlight=/I,%
linkcolor=rltred,%
citecolor=rltgreen,%
linktocpage=true}

\begin{document}
\setlength{\tabcolsep}{1pt}
\title{Percolating States in the Topological Anderson Insulator}
\author{Adrian Girschik}
\affiliation{Institute for Theoretical Physics, Vienna University of Technology, A-1040, Vienna, Austria, EU}
\author{Florian Libisch}
\affiliation{Institute for Theoretical Physics, Vienna University of Technology, A-1040, Vienna, Austria, EU}
\author{Stefan Rotter}
\affiliation{Institute for Theoretical Physics, Vienna University of Technology, A-1040, Vienna, Austria, EU}
\begin{abstract}
We investigate the presence of percolating states in disordered two-dimensional topological insulators.
In particular, we uncover a close connection between these states and the so-called topological Anderson insulator (TAI),
which is a topologically  non-trivial phase induced by the presence of disorder.
The decay of this phase could previously be connected to a delocalization of bulk states with increasing disorder strength.
In this work we identify this delocalization to be the result of a percolation transition of states that circumnavigate the hills of the bulk disorder potential.
\end{abstract}
\maketitle
\section*{Introduction}
Topological insulators are recently discovered materials with promising electronic properties \cite{HasKan2010,QiZha2011,And2013}.
The first physical system that was identified as a topological insulator was the HgTe/CdTe quantum well \cite{BerHugZha2006} featuring one-dimensional edge states
that are protected from backscattering by time-reversal symmetry, strongly stabilizing them against non-magnetic disorder.
The existence of such conducting edge states is hallmarked by a quantized conductance plateau that has meanwhile been verified experimentally. \cite{KoeWieBru2007}
As discovered in a successive numerical study, \cite{LiChuJai2009} such edge states are not only immune from backscattering
but can even be elicited by disorder in systems that have no topologically distinct properties in the clean limit.
This disorder-induced topological phase was first believed to be caused by Anderson localization, and was thus named ''topological Anderson insulator'' (TAI).
A later study found, however,
that the phase boundaries at the transition from an ordinary insulator to the TAI can be explained by an effective medium theory
in which the presence of disorder leads to a re-normalization of the medium parameters.\cite{GroWimAkh2009}
In this sense, the TAI appears due to a change of topology in the effective medium.

While the transition from an ordinary insulator to the TAI could be explained by the aforementioned effective description, \cite{GroWimAkh2009}
the transition from the TAI phase back to an ordinary insulating phase at very strong disorder values proves more involved:
the bulk states localize at intermediate disorder strength allowing for unimpeded edge-transport in the TAI phase, yet delocalize when disorder becomes even stronger. \cite{GroWimAkh2009,XinZhaWan2011,Pro2011}
So far, the resulting breakdown of the TAI phase could be attributed to the coupling of counter-propagating edge states on opposing edges through these delocalized bulk states,
resulting in a suppression of the edge states' immunity from backscattering. \cite{JiaWanSun2009,XuQiLiu2012}
This mechanism is responsible for an increased sensitivity to finite size effects \cite{LiZanJia2011,CheLiuLin2012} making the transition hard to explore numerically
and leaving the true nature of this counter-intuitive delocalization unclear.
While first studies \cite{CheLiuLin2012,Pro2011,YamNomImu2011} interpreted the bulk delocalization as an intermediate metallic phase,
a later study \cite{XuQiLiu2012} considering larger systems pointed out that only a single bulk state is probably responsible for the delocalization and an intermediate metallic phase is not present.
In addition, a spatially correlated potential and the associated pronounced bulk delocalization turn out to destroy the TAI phase entirely. \cite{GirLibRot2013}
In this work, we resolve the puzzle associated with these different observations by identifying the emergence of percolating states as the origin of the delocalization and
by clarifying the general connection between such states and the TAI phase.

\section*{Model and Methods}
As a starting point we choose the well-studied disordered HgTe/CdTe quantum well as described by the Bernevig-Hughes-Zhang (BHZ) model \cite{BerHugZha2006}
in terms of an effective 4-band Hamiltonian
\begin{equation} \label{effective hamiltonian}
H_{\rm{eff}} (k_x,k_y)=
\left(
\begin{array}{cc}
h(\vec k) & 0 \\
0 & h^*(-\vec k)
\end{array} 
\right) 
\end{equation}
with 
\begin{align}
h(\vec k)&=\mathds{1} \epsilon(\vec k) + d_i(\vec k) \sigma^i \label{h(k)} \\
\epsilon(\vec k)&=C-D\left(k_x^2+k_y^2\right)  \nonumber  \\
d_i&=\left( \begin{array}{ccc} Ak_x,& -Ak_y,& M(\vec k) \end{array}\right)^T \nonumber \\
M(\vec k) &= m-B(k_x^2 + k_y^2) \nonumber
\end{align}
and $\sigma^i$ representing the Pauli-matrices.
Following Ref.~\onlinecite{KoeBuhMol2008}, we choose the following set of realistic quantum well parameters in all our computations:
$A=364.5\;\rm{nm}\cdot\rm{meV}$, $B=-686.0\;\rm{nm}^2\cdot\rm{meV}$ and $D=-512.0\;\rm{nm}^2\cdot\rm{meV}$.
The topology of the system is determined by the sign of the topological mass $m$:
For $m<0$ the bulk band gap of size $2|m|$ is topologically non-trivial and thus filled with gap-less edge states characterizing a two-dimensional topological insulator.
On the other hand, for $m>0$, the bulk band gap is topologically trivial and does not contain any states leaving us with a system that is an ordinary insulator.

\begin{figure} 
    \begin{center}
        \includegraphics[angle=0, scale=1.0, width=0.45\textwidth]{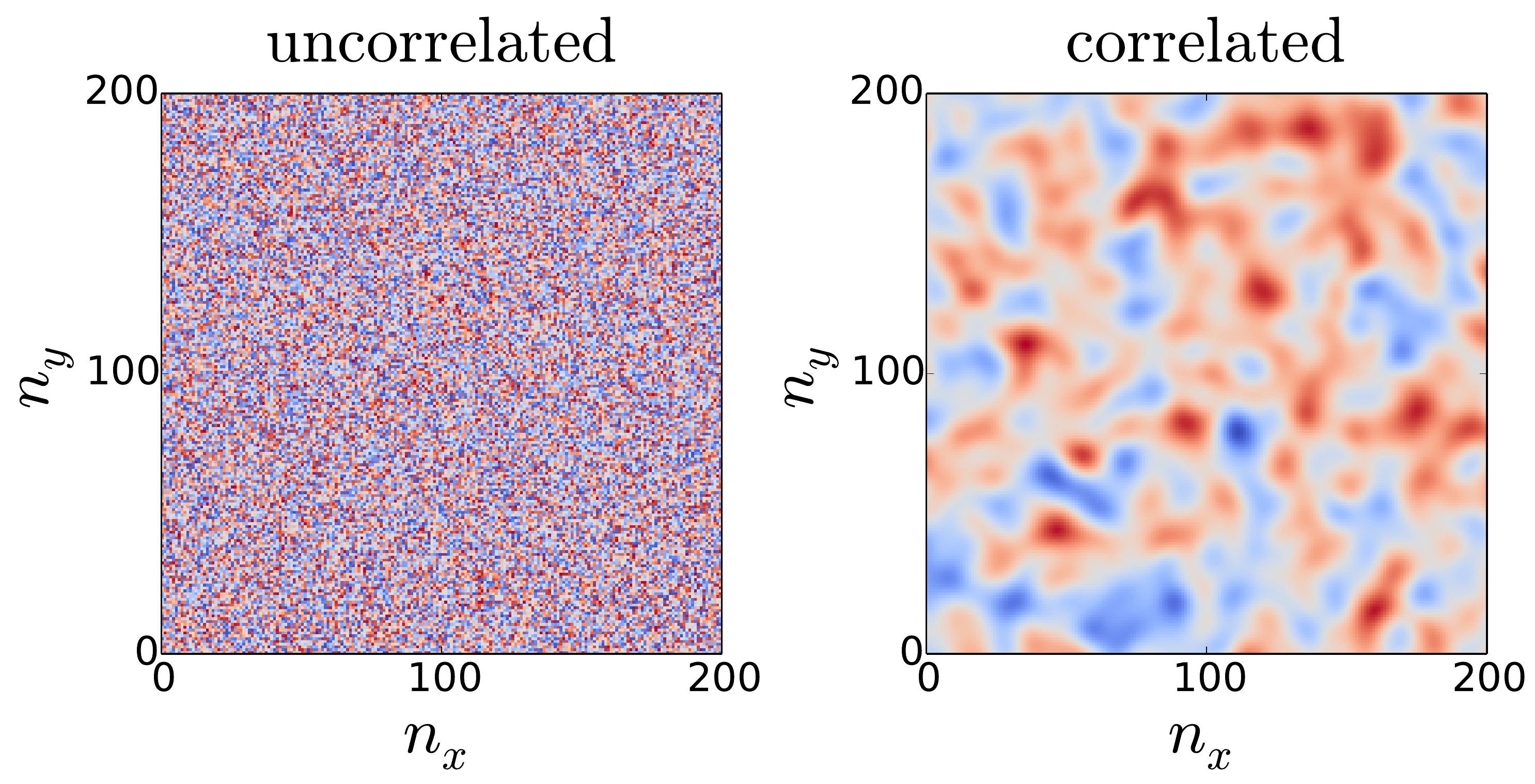} 
        \caption{
            (Color online) Sketch of an uncorrelated disorder potential in a system of width and length $W=L=1000\,\rm{nm}$ and a correlated potential with correlation length $\xi=35\,\rm{nm}$.
            The color code ranges from blue (strong negative potential) to red (strong positive potential).
        \label{fig:pot_sketch}}
    \end{center}
\end{figure}

Using the advanced modular recursive Green's function method \cite{RotWeiRoh2003,RotTanWir2000,LibRotBur2012}
we calculate the conductance $G$ through two-dimensional rectangular ribbons of HgTe/CdTe quantum wells discretized on a square grid with discretization constants $\Delta x$ and $\Delta y$,
width $W=(n_y+1) \cdot \Delta y$ and length $L=n_x\cdot\Delta x$. 
In accordance with previous studies, \cite{LiChuJai2009,JiaWanSun2009,GroWimAkh2009,LiZanJia2011,Pro2011,XuQiLiu2012,CheLiuLin2012,ZhaShe2013,GirLibRot2013}
the discretization constants are set to $\Delta x =\Delta y = a = 5\;\rm{nm}$.
Two clean semi-infinite leads are attached to the left and right end of the ribbon.
Following the Landauer-B\"uttiker formalism, the conductance $G$ in the limit of vanishing temperature is given by the total transmission $T$ at the Fermi energy $E_F$.
Our method also allows for a calculation of the scattering wave functions $\psi(x,y)$ as well as the density of states $\rho(E) = -{\rm Tr}\left[ {\rm Im}\, G^r(\vec{x},\vec{x},E)\right]/\pi$
where $G^r$ is the retarded Green's function.
Disorder is modelled by static on-site energy values $V(x_i,y_i)$ at each grid point $(x_i,y_i)$ randomly chosen from the interval $[-U/2,U/2]$ with $U$ the disorder strength.
In most studies, \cite{LiChuJai2009,JiaWanSun2009,GroWimAkh2009,LiZanJia2011,Pro2011,XuQiLiu2012,CheLiuLin2012,ZhaShe2013} the values of $V(x_i,y_i)$ are chosen without any spatial correlations
between neighboring grid points [see Fig.~\ref{fig:pot_sketch}a].
Here, we also consider spatial correlations in $V(x,y)$ [see Fig.~\ref{fig:pot_sketch}], characterized by a finite correlation length $\xi$, which
can significantly affect the conduction properties.\cite{GirLibRot2013}

\section*{Results}
\begin{figure} 
    \begin{center}
        \includegraphics[angle=0, scale=1.0, width=0.45\textwidth]{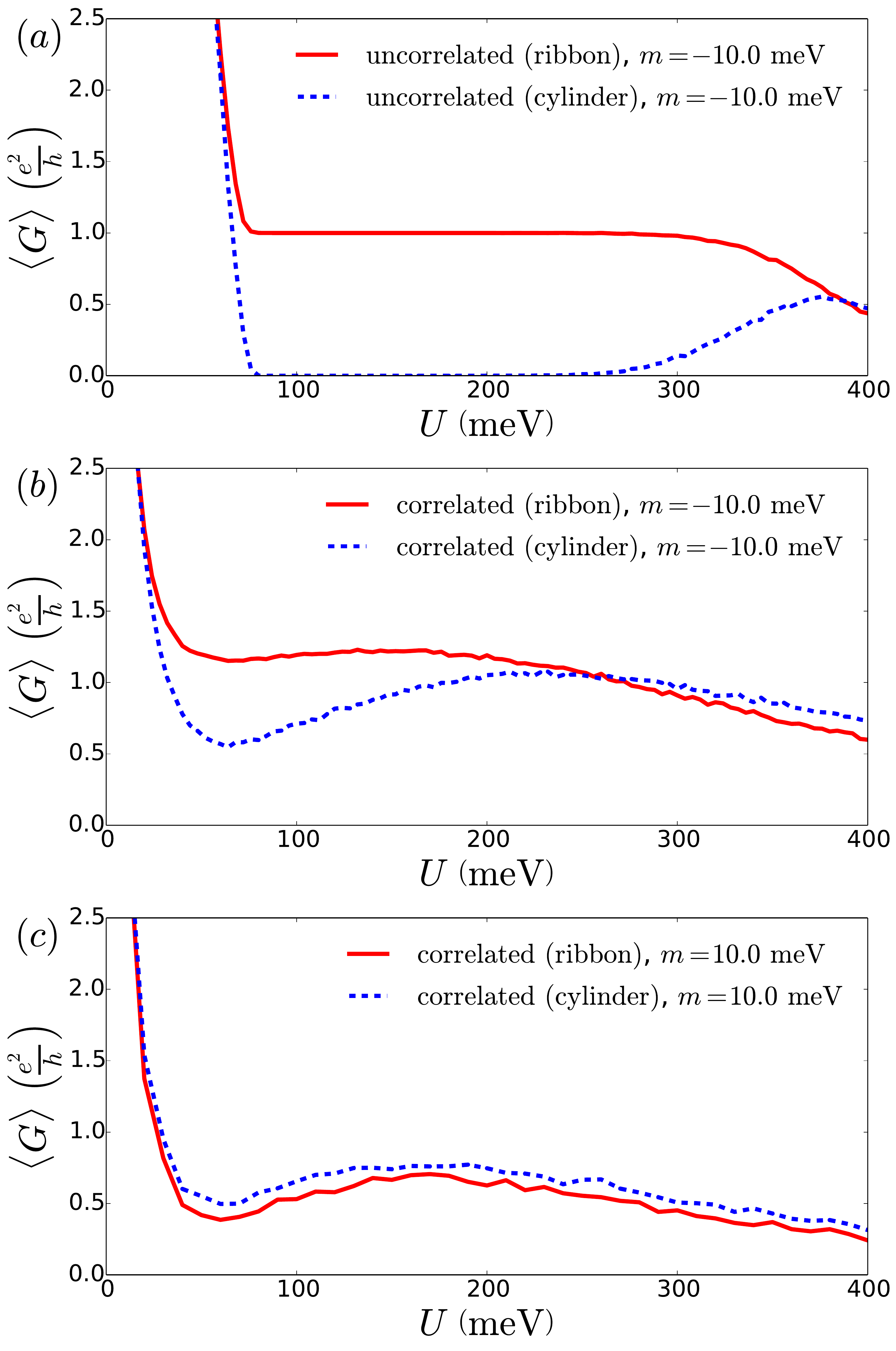} 
        \caption{
            (Color online) Average conductance $\langle G\rangle$ as a function of disorder strength $U$ through different HgTe/CdTe quantum wells.
            All systems are of width and length $W=L=1000\,\rm{nm}$ and the Fermi energy is set to $E_F=16\,\rm{meV}$.
            The ribbon geometries are shown by red solid curves while the cylinder geometries are shown by the blue dashed lines.
            (a) Average conductance through a system of topological mass $m=-10\,\rm{meV}$ (topological insulator) in an uncorrelated potential.
            The disorder average is taken over $1000$ configurations.
            The ribbon geometry features the TAI conductance plateau of $\langle G\rangle=1 e^2/h$ for $80\,\rm{meV}\lesssim U \lesssim 280\,\rm{meV}$.
            In this region the conductance through the cylinder is almost entirely suppressed followed by a delocalization transition of the bulk states starting at $U\approx 280\,\rm{meV}$.
            (b) Average conductance through systems with the same parameters as in (a) but in a correlated potential with correlation length $\xi=35\,\rm{nm}$.
            The cylinder geometry (blue dashed curve) clearly shows the bulk delocalization transition, which occurs here for weaker disorder strengths $U$ than in the uncorrelated case.
            (c) Average conductance through systems with positive topological mass $m=10\,\rm{meV}$ (ordinary insulator) in a correlated potential of correlation length $\xi=35\,\rm{nm}$.
            The disorder average is taken over $200$ configurations.
            The cylinder geometry shows a less pronounced bulk delocalization transition than in (b), followed here also by the results for the ribbon due to the absence of edge states.
        \label{fig:transmissions} }
    \end{center}
\end{figure}

In our simulations we first consider a quadratic region of width $W=L=1000\,\rm{nm}$ and an uncorrelated disorder potential (i.e., $\xi\to 0$).
Two geometries will be studied that only differ in their boundary conditions:
a ribbon for which hard wall boundary conditions along the edges are applied and a cylinder with periodic boundary conditions in $y$-direction.
A comparison between the disorder-averaged conductance $\langle G\rangle$ through these two geometries has been used previously to distinguish between bulk
and edge phenomena as the periodicity of the cylinder eliminates the edges of the geometry. \cite{JiaWanSun2009}
The results for $\langle G\rangle$ in a topological insulator with $m=-10\,\rm{meV}$ at Fermi energy $E_F=16\,\rm{meV}$ are shown in Fig.~\ref{fig:transmissions}(a)
as a function of the disorder strength $U$ for both geometries.
The value of $E_F$ is chosen such that for uncorrelated disorder the TAI conductance plateau with $\langle G \rangle = 1\, e^2/h$ clearly appears
in the ribbon \cite{LiZanJia2011,JiaWanSun2009,XinZhaWan2011,XuQiLiu2012} for disorder strength $80\,\rm{meV}\le U\le 280\,\rm{meV}$ [see red curve in Fig.~\ref{fig:transmissions}(a)].
In the cylinder this plateau is clearly absent, since no edge states can exist in this edge-less geometry.
While for disorder values beyond this plateau the conductance drops monotonically in the ribbon,
the conductance through the cylinder geometry [blue dashed curve in Fig.~\ref{fig:transmissions}(a)] shows a renewed increase at the same disorder strength.
This is the signature of the aforementioned bulk delocalization that has already been observed in uncorrelated potentials. \cite{GroWimAkh2009,XinZhaWan2011,Pro2011}
A physical intuition for this transition is, however, still lacking, but will become clear by considering disorder potentials
that are spatially correlated [illustrations for uncorrelated and correlated disorder potentials are shown in Fig.~\ref{fig:pot_sketch}].

In a previous work \cite{GirLibRot2013} we demonstrated that spatial correlation in the disorder potential can destroy the TAI phase entirely.
Here we consider a situation, in which the correlations all but dissolve the plateau in the ribbon geometry [see the red curve in Fig.~\ref{fig:transmissions}(b)].
For these parameter values it is best visible that the dissolution of the plateau is accompanied by a delocalization of bulk states.
As can be seen by comparing the blue dashed curves in Fig.~\ref{fig:transmissions}(a) and (b),
this delocalization happens at much lower disorder values for correlated potentials than for uncorrelated ones.
In both cases, however, these delocalized bulk states contribute to the conductance, but also suppress the edge conductance by coupling counter-propagating edge states with each other,
thereby leading to a breakdown of the TAI conductance plateau.

\begin{figure} 
    \begin{center}
        \includegraphics[angle=0, scale=1.0, width=0.45\textwidth]{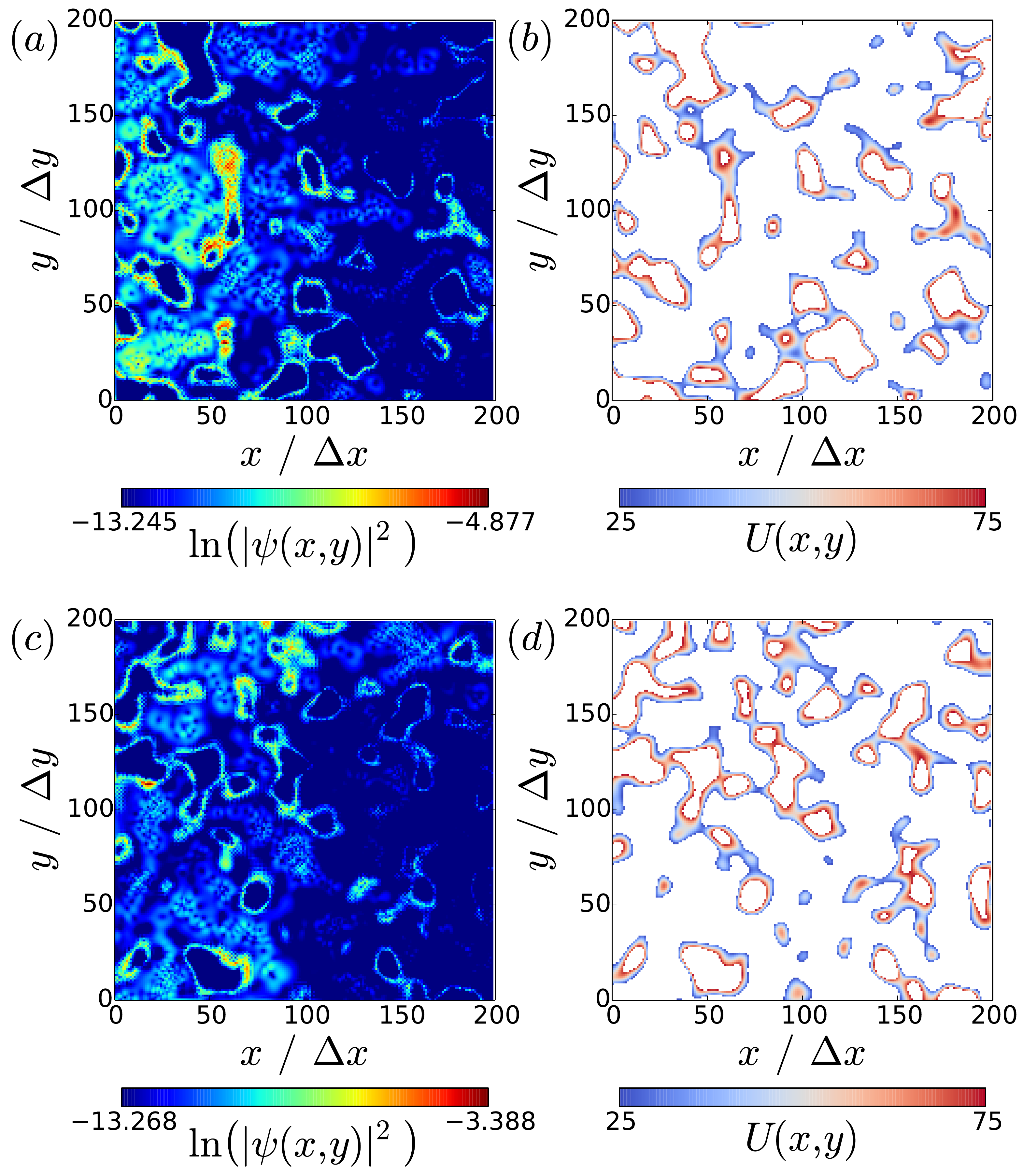} 
        \caption{(Color online) Comparison of the scattering wave function $|\psi(x_i,y_i)|^2$ (shown on a logarithmic scale in the left column)
            and the potential itself (right column) at the delocalization transition ($U=220\,\rm{meV}$) for systems with $W=L=1000\,\rm{nm}$,
            $E_F=16\,\rm{meV}$ and $m=-10\,\rm{meV}$.
            We only colored the potential values between $26\,\rm{meV}\le V(x_i,y_i) \le 78\,\rm{meV}$ for which the Fermi energy is effectively shifted into the valence band
            of the clean band structure (the remaining potential values are left in white).
            Two different disorder configurations for cylinder and ribbon geometry are considered in the top and bottom row, respectively.
            The similarity of the wave functions and these truncated potentials illustrates that the delocalizing bulk states are percolating around the hills of the potential landscape and that
            these percolating states have their origin in the bulk band structure of the clean sample.
        \label{fig:wf_pot_compare}}
    \end{center}
\end{figure}

\begin{figure} 
    \begin{center}
        \includegraphics[angle=0, scale=1.0, width=0.45\textwidth]{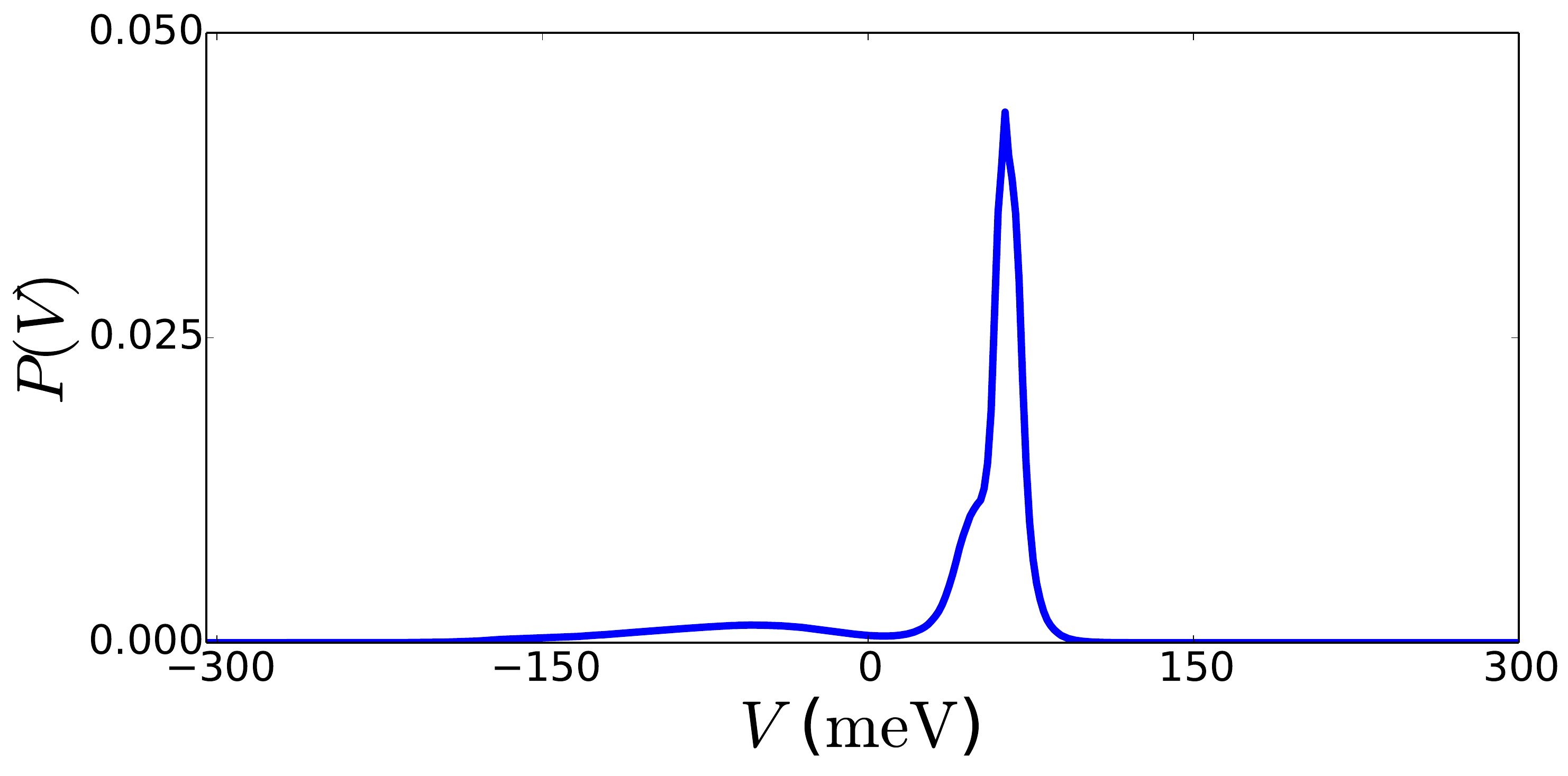} 
        \caption{
            (Color online) Probability density distribution $P(V)$ for disorder values $V(x_i,y_i)$ at grid points $(x_i,y_i)$ weighted by the absolute value $|\psi(x_i,y_i)|^2$
            of the scattering wave function at $E_F=16\,\rm{meV}$ with injection in the first lead mode in a system of width and length $W=L=1000\,\rm{nm}$ and topological mass $m=-10\,\rm{meV}$.
            A disorder average is taken over $1000$ configurations of random potentials with disorder strength $U=220\,\rm{meV}$ and correlation length $\xi=35\,\rm{nm}$.
            The distribution shows an enhancement of the wave functions at disorder values $V$ situated between $V_{\rm{min}}\approx25\,\rm{meV}$ and $V_{\rm{max}}\approx75\,\rm{meV}$.
            These values correspond nicely to the distance of Fermi energy $E_F$ to the flat valence bulk band states at the BZ edges.
        \label{fig:single_histogram} }
    \end{center}
\end{figure}

To get a better insight into this scenario, we now consider the scattering wave functions $\psi(x_i,y_i)$
during this delocalization transition [see Fig.~\ref{fig:wf_pot_compare}(a) and (c) for two such states at $U=220\,\rm{meV}$].
A first optical inspection of these wave function images suggests that the associated flux is circumnavigating the hills of the underlying correlated potential, \cite{GirLibRot2013}
reminiscent of percolation states found in the Quantum Hall effect \cite{HasSohWie2008} and in antidot topological insulator lattices. \cite{ChuLuShe2012}
To make this first impression more quantitative, we analyze how the intensities of the wave functions $\psi(x_i,y_i)$
shown in  Fig.~\ref{fig:wf_pot_compare} are correlated with the values $V(x_i,y_i)$ of the underlying potential landscape.
For this purpose we compute the weighted probability $P(V)$ for a wave to encounter a potential value $V$ with the weights of this probability distribution
being given by the intensity $|\psi(x_i,y_i)|^2$ of the wave function at a grid-point $(x_i,y_i)$ with potential value $V$.
The distribution $P(V)$ resulting from an average over 1000 disorder realizations shows a surprisingly pronounced enhancement at positive disorder values $V$
approximately situated between $V_{\rm min}\approx 25\,\rm{meV}$ and $V_{\rm max}\approx 75\,\rm{meV}$ [see Fig.~\ref{fig:single_histogram}],
suggesting that disorder values from this interval give rise to clearly enhanced wave function intensities.
Apparently the states responsible for the bulk delocalization tend to reside primarily at relatively high values of the disorder potential, i.e.,
in a certain altitude interval of the hills in the correlated potential landscape.
Correspondingly, we find that the wave function intensities shown in Fig.~\ref{fig:wf_pot_compare}(a) and (c) strongly resemble contour plots of the associated disorder potential,
when we truncate that latter to the interval $V\in[25,75]$ meV [see Fig.~\ref{fig:wf_pot_compare}(b) and (d)].

\begin{figure} 
    \begin{center}
        \includegraphics[angle=0, scale=1.0, width=0.45\textwidth]{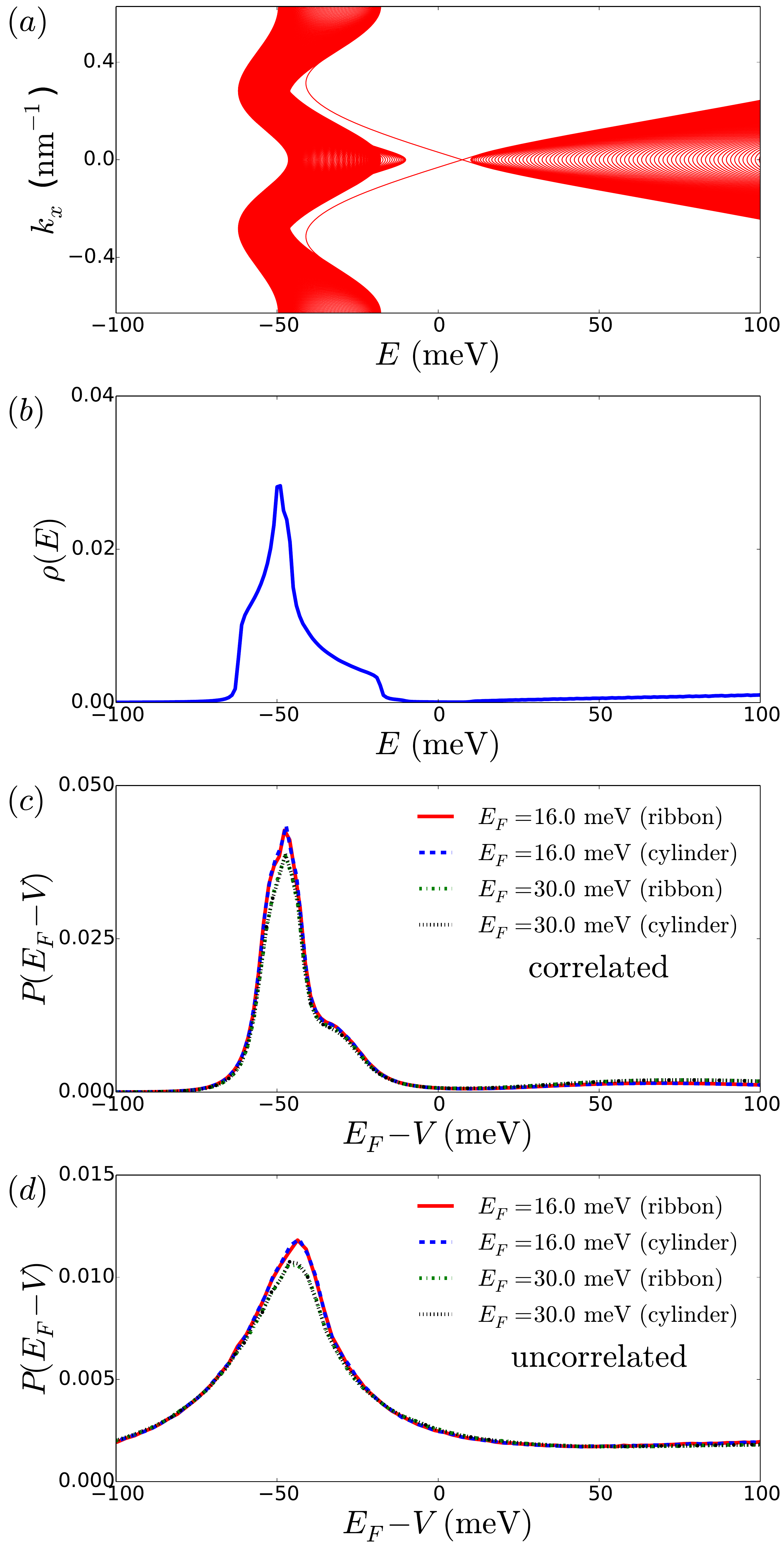} 
        \caption{
            (Color online) (a) Band structure $E(k_x)$ of a system of width $W=1000\,\rm{nm}$ and topological mass $m=-10\,\rm{meV}$ obtained by directly solving the Bloch-eigenvalue problem
            numerically for the Hamiltonian in Eq.~(\ref{effective hamiltonian}).
            (b) Density of states $\rho(E)$ in a closed and clean system of width and length $W=L=1000\,\rm{nm}$, topological mass $m=-10\,\rm{meV}$ and cylindrical geometry.
            (c) Weighted probability density distribution $P(E_F-V)$ for systems with $m=-10\,\rm{meV}$, $W=L=1000\,\rm{nm}$ in a correlated potential of correlation length $\xi=35\,\rm{nm}$.
            The wave functions are calculated in the region of the bulk delocalization at disorder strength $U=220\,\rm{meV}$ and for $1000$ random disorder configurations.
            The red solid and blue dashed lines show $P(E_F-V)$ at $E_F=16\,\rm{meV}$ in the ribbon and the cylinder geometry.
            The green dash-dotted and black dotted lines represent $P(E_F-V)$ for ribbon and cylinder at $E_F=30\,\rm{meV}$.
            The similarity of all four curves shows that the distribution is a very fundamental system property.
            The shape of the distributions closely resembles the density of states at the BZ boundary shown in (b).
            This indicates the bulk-like nature of the states responsible for the delocalization transition.
            (d) Weighted probability density distribution $P(E_F-V)$ with the same system parameters as in (c) but in an uncorrelated disorder potential.
            The pictures are again taken in the region of the bulk delocalization at disorder strength $U=370\,\rm{meV}$ and for $1000$ disorder configurations.
        \label{fig:histogram_figure}}
    \end{center}
\end{figure}

To identify the origin of this curious behavior, we first point to the fact that the above interval bounds, i.e., $V_{\rm min}\approx 25\,\rm{meV}$ and $V_{\rm max}\approx 75\,\rm{meV}$,
are astonishingly close to the minimal and maximal distances $E_{\rm{min}}=26\,\rm{meV}$ and $E_{\rm{max}}=78\,\rm{meV}$ of the Fermi energy $E_F=16\,\rm{meV}$ to the energy range
$E_{\rm{bulk}}=[-10,-62]\,\rm{meV}$ in which the valence band states are situated in the clean system [see the band structure of Fig.~\ref{fig:histogram_figure}(a)].
This observation suggests that the flux in our correlated potential is carried mostly by the disorder-analogues of these valence band states.
Further evidence for this correspondence can be deduced when considering the rescaled probability distribution $P(E_F-V)$, which measures, as above,
the probability for a scattering state to reside at a potential value $V$, but now shifted by the Fermi energy $E_F$.
We find that this distribution, quite remarkably, stays almost invariant with respect to a change of the Fermi energy
[see a comparison between two different values of $E_F$ in Fig.~\ref{fig:histogram_figure}(c)].
This observation reflects the fact that a change of $E_F$ just shifts the corresponding wave functions to different disorder values,
but that the origin of states in the valence bands stays unchanged.
Furthermore, the density of states $\rho(E)$, shown in Fig.~\ref{fig:histogram_figure}(b), and the distribution $P(E_F-V)$, shown in Fig.~\ref{fig:histogram_figure}(c), are very similar -
even the small kinks in $\rho(E)$ are clearly reproduced in $P(E_F-V)$.
Such kinks in the density of states are nothing else but van-Hove-singularities resulting from the flat bands in the band structure.
We may thus conclude that the flat valence band states especially at the Brillouin zone (BZ) edges (where the maximum of $P(E_F-V)$ is located) represent the most significant contribution
to the intensity of the scattering wave functions.
In addition, we find that the distribution $P(E_F-V)$ is not at all sensitive to the boundary conditions since it is almost exactly equal for the ribbon and for the cylinder
(see Fig.~\ref{fig:histogram_figure}(b)).
The distribution $P(E_F-V)$ thus turns out to be quite fundamental in that it has its origin in basic system properties,
which are given here by the band structure in Fig.~\ref{fig:histogram_figure}(a) and by the flat band states contained in it.

These observations allow us to construct a comprehensive picture of the physics in the cylindrical geometry with a correlated potential [see blue dashed curve in Fig.~\ref{fig:transmissions}(b)]:
While in the clean limit pure bulk conduction is observed,
the conductance drops down to a minimum at disorder strength of $U\approx 60\,\rm{meV}$ due to the increasing localization of the bulk states [see Fig.~\ref{fig:transmissions}(b)].
When at $U\approx 100\,\rm{meV}$ the hills of the correlated potential are high enough to locally shift the Fermi energy into the valence band,
the bulk is filled with localized states deriving from clean valence band bulk states spatially located around the hills of the underlying disorder potential.
With growing disorder strength, these localized states connect with each other and go through a percolation threshold,
which is  responsible for the delocalization transition and the increased bulk conductance.
Only at very strong disorder $U\approx 300\,\rm{meV}$ the connection between these percolating states weakens and the conductance again decreases.
This picture is also strongly supported by previous studies of the TAI in the uncorrelated case (see Refs.~\onlinecite{ZhaChuZha2012} and \onlinecite{ZhaShe2013}):
Considering the arithmetic and geometric average of the local density of states it was shown there
that the states carrying the flux in the TAI are not single extended states throughout the whole TAI phase (as would be expected for edge states)
but for very strong disorder rather formed by clusters of well localized states.
Our percolating wave functions deriving from the valence band are perfect candidates for such linked, localized states.
This picture is also corroborated by the flatness of the valence band states,
which leads to the very small group velocity responsible for the wave function enhancements around the potential hills
as seen in the examples of Fig.~\ref{fig:wf_pot_compare}(a) and (c).

The flatness of the states in the effective band structure is, in fact, also important for the theory put forward in the aforementioned studies: \cite{ZhaChuZha2012,ZhaShe2013}
Considering disordered super-cell structures it was argued that flat and localized bands develop small sub-gaps that can be of topological non-trivial type.
Hence these gaps have to be filled with edge states in the same way as the inverted band gap of a clean topological insulator is. \cite{BerHugZha2006,KoeWieBru2007,KoeBuhMol2008}
In this picture the TAI phase is thus characterized by edge states that appear in the energy gaps between localized bulk states and are consequently again immune from backscattering.

At this point the question arises how the above results can be reconciled with our own model, which so far does not contain any reference to edge states in the percolation transition.
To investigate this issue in detail, we performed additional calculations for a system where no edge effects can be present due to a topological mass,
which we choose to take on the positive value of $m=10.0\,\rm{meV}$.  As shown in Fig.~\ref{fig:transmissions}(c) this sign change of $m$ significantly modifies the conductance properties.
While previously for $m<0$ and moderate disorder strength $U$ the conductance in the ribbon was clearly enhanced in comparison to the cylinder [see Fig.~\ref{fig:transmissions}(b)],
the conductance of the ribbon for $m>0\,$ is even smaller than in the cylinder [see Fig.~\ref{fig:transmissions}(c)].
This behavior can be attributed to the absence of edge states at the sample edges for positive topological mass $m>0$.
In the cylindrical geometries we find that the delocalization transition is less pronounced for $m>0$ than it was for $m<0$ [compare blue dashed lines in Figs.~\ref{fig:transmissions}(b) and (c)].
On the one hand the fact that the delocalization transition still exists for $m>0$ supports our model of flat bulk states undergoing a percolation transition.
On the other hand, however, the more pronounced nature of the transition for $m<0$ suggests
that edge states propagating along the edges of the potential hills provide an additional link between localized states leading to a larger conductance.
This picture, indeed, agrees very well with the analyses of Refs.~\onlinecite{ZhaChuZha2012} and \onlinecite{ZhaShe2013},
since the connecting local edge states in our model can directly be identified with the edge states that were predicted to form in the non-trivial sub-gaps of the localized flat bands.

We would thus be in a perfect position to complement the theory of Refs.~\onlinecite{ZhaChuZha2012,ZhaShe2013} with the intuitive explanation that these sub-gap edge states exist locally and 
connect bulk states localized around hills of the potential landscape to form a percolating network of internal bulk states that 
lead to the decay of the TAI phase. The missing piece to complete our argument is to show that the picture we derived for the case of correlated disorder holds also for the uncorrelated case 
considered in Refs.~\onlinecite{ZhaChuZha2012,ZhaShe2013}.

We check this point explicitly, by verifying that our model can explain the appearance of the TAI as well as the observed delocalization-localization transition of the bulk states
for the case of uncorrelated disorder.
Consider, in this context, that the TAI conductance plateau in the ribbon geometry between $U\approx \,80\rm{meV}$ and $U\approx 280 \,\rm{meV}$ [see red curve in Fig.~\ref{fig:transmissions}(a)]
is destroyed by the onset of the bulk delocalization at $U\approx 280\,\rm{meV}$ in the cylinder [see blue dashed curve Fig.~\ref{fig:transmissions}(a)] which happens for much larger $U$ than
in the correlated case.
Still, when the delocalization transition is in full effect (at $U= 370 \,\rm{meV}$)
the corresponding scattering states show a similar weighted distribution $P(E_F-V)$ in the now spatially uncorrelated potential
as already observed in the correlated case [see Fig.~\ref{fig:histogram_figure}(d)].
Again the peak of this distribution fits nicely to the band structure of the clean limit,
indicating that our picture of local edge states percolating around internal edges of strong disorder holds also for the uncorrelated case.
Last but not least, we mention that such a percolating state corresponds exactly to the ''single bulk state'' that is held responsible for the delocalization in Ref.~\onlinecite{XuQiLiu2012}.

\section*{Discussion}
Our results suggest that the emergence as well as the decay of the TAI phase depend strongly on the energy offset and on the flatness of bulk valence bands in the clean limit.
These flat bands feature an enhanced contribution to the density of states and occur in the center as well as at the boundaries of the BZ.
Yet, the underlying BHZ model is only valid for small $k_x$ close to the $\Gamma$-point and thus does not yield a good approximation for the valence bands at the BZ boundaries of a real HgTe/CdTe
quantum well. \cite{BerHugZha2006}
Correspondingly, we find that when changing the grid spacing $a$ in our discretized lattice from the value conventionally used in the literature ($a=5\,\rm{nm}$) to different values,
the position of the BZ boundaries $k_x^{\rm{BZ}}=\pm \pi / a$ and the energy offset of these states at the BZ boundaries also change significantly.
We also verified that the flatness of the bands at the BZ boundary is a direct consequence of the discreteness of the underlying lattice used for the numerical solution of the transport problem
(see also Refs. \onlinecite{LiChuJai2009,JiaWanSun2009,GroWimAkh2009} where discretized models were first employed to describe the TAI).
As a result, the localization-delocalization transition and possibly even the TAI phase itself associated with these states at the BZ boundary will not occur in real HgTe/CdTe quantum wells
as the strong-disorder limit in these devices will be different from the predictions of the discretized model.
Quite remarkably, however, realizations of topological insulators have recently also been considered based on photonic systems. \cite{RecZeuPlo2013,TitLinRec2015}
These so-called Floquet topological insulators are based on a discretized lattice of sites, just like in the numerical model used above to approximate the physics in HgTe/CdTe quantum wells.
The strong-disorder physics, which we have discussed here, may thus well be realized in experiments based on effective model systems in optics \cite{RecZeuPlo2013,TitLinRec2015} as well as in
acoustics \cite{YanGaoShi2015} or in other fields where wave scattering parameters can be tuned appropriately.

\section*{Conclusion}
In this work we uncover the existence of percolating states in two-dimensional topological insulators.
In particular, we show how these states affect the phase boundaries of the topological Anderson insulator,  which is a topologically non-trivial phase caused by disorder.
While the reason for the emergence of this phase has already been understood, its breakdown could so far only be vaguely connected to a delocalization of bulk states.
Here we show that in a spatially correlated potential this delocalization is caused primarily by  bulk states, that are localized when circumnavigating the hills of the disorder potential,
but that become connected with each other when passing a percolation threshold.
These connections and thus also the delocalization transition are consolidated by local edge states that can internally form in the disordered sample.
By showing how the localized bulk states derive from flat bands in the valence band structure of the clean sample without disorder,
we clarify that the same physics is at work also in the well-studied case of an uncorrelated disorder potential.
\clearpage
\pagebreak

\clearpage
\bibliography{TIpaper2}
\end{document}